\documentclass[a4paper]{revtex4}
\usepackage{inputenc}
\usepackage[english]{babel}

\usepackage{array,booktabs}
\usepackage{array} 
\usepackage{lipsum}   
\usepackage{calc}
\RequirePackage{graphicx}
\usepackage{color}
\usepackage{amsmath}
\usepackage{float}
\usepackage{calc}
\usepackage{pdflscape}
\usepackage{color}
\usepackage{float}
\usepackage{subfigure}
\usepackage[font=small,labelfont=bf]{caption}
\usepackage{graphicx}
\usepackage{pdflscape}
\usepackage{color}
\usepackage{float}
\usepackage[font=small,labelfont=bf]{caption}
\usepackage{graphicx}
\usepackage{amsmath}
\usepackage[nodisplayskipstretch]{setspace}

\usepackage{setspace}
\usepackage{tabularx}

\usepackage{float}
\usepackage{color}
\usepackage{amsmath}
\usepackage{float}
\usepackage{calc}
\usepackage{pdflscape}
\usepackage{color}
\usepackage{float}
\usepackage{subfigure}
\usepackage[font=small,labelfont=bf]{caption}
\usepackage{graphicx}
\begin{document}{\setlength\abovedisplayskip{4pt}}
\title{Majorana Neutrinos and Clockworked Yukawa Couplings contribution to non-observation of the rare leptonic decay $ l_{i}\rightarrow l_{j} \gamma $, Clockwork Photon and Clockwork Graviton.}

\author{Gayatri Ghosh}
\email{gayatrighdh@gmail.com}
\affiliation{Department of Physics, Gauhati University, Jalukbari, Assam-781015, India}
\affiliation{Department of Physics, Pandit Deendayal Upadhayay Mahavidyalaya, Karimganj, Assam-788720, India}
\begin{abstract}
The clockwork is an extra-dimensional set-up for generating light particles with exponentially suppressed or hierarchical couplings of light particles with N massive states having comparable masses near the threshold scale of the mechanism in theories which contain no small parameters at the fundamental level. We explore the prospect of charged lepton flavour violation (cLFV) in a clockwork framework which encompasses Dirac mass terms as well as Majorana mass terms for the new clockwork fermions. We deive the masses of the non zero clockwork Majorana masses, and new particles in a clockwork framework and for their Yukawa couplings to the lepton doublets, in the framework where the clockwork parameters are universal. When the new clockwork Majorana masses are non zero, neutrino masses are generated as a result from the exchange of heavy messenger particles such as right handed iso$-$singlet neutrinos or iso$ - $triplet scalar bosons known as the seesaw mechanism. In the case of non zero clockwork Majorana masses, owing to the sizable effective Yukawa couplings of the higher mass modes neutrino masses can only be made tiny by conjecturing large Majorana mass in the teraelectron volt range for all
the clockwork gears. This is apparent from the constraints on the mass scale of the clockwork fermions due to the non-observation of the rare cLFV decay $ \mu\rightarrow e\gamma $, $ \tau \rightarrow \mu \gamma $, $ \tau\rightarrow e\gamma $. A general description of the clockwork mechanism valid for fermions, gauge bosons, and gravitons is discussed here. This mechanism can be implemented with a discrete set of new fields or, in its continuum version, through an extra spatial dimension. In both cases the clockwork emerges as a useful tool for model-building applications. Notably, the continuum clockwork offers a solution to the Higgs naturalness problem, which turns out to be the same as in linear dilaton duals of Little String Theory. 
\end{abstract}   
\maketitle
\section{Introduction}
Beyond Standard Model Physics can be explained at low energy as non renormalisable field operators involving Standard Model fields. There is a great syndicate between the interaction scales and mass scales like the Weinberg operator which may be generated via two right handed neutrinos oppositely charged under $U(1)^{'}$. Interaction scales and mass scales are totally distinct quantities. When measured in natural units, the difference betweeen mass and scales has little fall out if the couplings are of the order  1. But the dynamics gets totally changed in presence of tiny yukawa couplings {\color{blue}\cite{q,w,e}} which are small in natural units since now the interaction scales enter low scale energies. In this work we discuss lepton flavor violation in a generalisation of clockwork models where large effective interaction scales or full UV completion enters at energies exponentially smaller than intimated by a given interaction strength .
\par
Neutrino's tiny mass is one of the most challenging open questions in Fundamental Physics and points to new laws of particle physics and reshape theories of cosmology. One probable solution to this enigma is provided by the seesaw mechanism, in which the smallness
of neutrino masses is unravelled by the violation of the lepton number at a very high energy scale by two units $ \Delta_{L} =2$ {\color{blue}\cite{a,b,c,d}}. Tiny yukawa couplings of the neutrino  on the other hand, couples to the Standard Model Higgs in models with lepton number conservation.  Occurrence of tiny Yukawa couplings is a
phenomenologically feasible possibility, and can be worked out in further extensions of the model. A new mechanism of producing small couplings in theories coupled to the Standard Model has been introduced {\color{blue}\cite{q,w,e,8,9,10,11,12,13,14,15}}. The implicative mechanism, of deconstruction models, is a linear model with no large hierarchies in the site dependent field theory parameters, which essentially arouses site-dependent suppressed couplings to the zero$-$mode {\color{blue}\cite{16}}. Originally, the clockwork theory was proposed for a quiver of axions, it is now generalized to spin-0 scalar, spin-$ \frac{1}{2} $ fermion, spin-1 boson, or spin-2 graviton and other fields {\color{blue}\cite{17}}. A clockwork fermion elucidates a very small Dirac neutrino mass or explain the hierarchical pattern of quark and lepton masses. With a clockwork gauge boson the presence of tiny gauge charges is justified. Clockwork graviton, solves the naturalness problem of the electroweak scale, syndicating an effective explanation for the weakness of gravity. Independently of the specific implementation of the clockwork, the theory proposes the existence of $N$ particles at the univesal mass scale $m$, which are known as the ‘clockwork gears’, as they are the degrees of freedom. The gears have universal properties and their exploration of the clockwork will also lead us to consider the limit $N$ tends to  infinity eventually the continuum limit, in which the 1$-$D space lattice in field space is elucidated as a physical spatial dimension. Some studies on the applications and generalizations of clockwork mechanism is done {\color{blue}\cite{18,19,20,21,22,23,24,25,26,27,28,29,30,31,32,33,34}}.
\par
In this work we explore the impact of Majorana neutrinos in case of charged lepton flavour violation in the fermionic clockwork model and the constraints on the masses of the clockwork gears. Concretely, we identify the  masses of the clockwork gears with the zero modes of a clockwork sector, which fits the neutrino masses and charged lepton masses, such that tiny yukawa couplings are naturally generated and therefore tiny neutrino masses. We establish the clockwork framework for the right handed neutrinos by counting  Majorana mass terms. We show how in the clockwork theory the suppression of the Yukawa couplings by field dependent exponential factors, is not altered by the manifestation of the Majorana mass terms. In fact, the combination of
the clockwork “suppression” and the Majorana “seesaw” sets now the mass constraints on the clockwork gears. The clockwork mechanism suppresses the the zero mode couplings, the yukawa couplings of the higher modes induce, via loops, potentially large sizable rates for the leptonic rare decays. The rest of the paper is organized as follows. In section 2, we present the most general phenomenology for clockwork neutrinos with Dirac and Majorana mass terms. In section 3, we discuss the impact of Majorana neutrinos on lepton flavour violation in the clockwork scenario and calculate constraints on the clockwork gear masses. We close with a summary.
\section{Clockwork Fermion}
Fermions may be massless due to a chiral symmetry. Fermion realisation of the
clockwork model requires a single chiral symmetry which is splitted amongst a number of field site in the underlying model. The left out chiral symmetry drags the massless
fermion exponentially to one edge of the clockwork. To this climax, one introduces $N +1$ chiral fermions $\psi_{Rj} (j = 0, . . . , N )$ together with N fermions $\psi_{Li}(i = 0, . . . , N-1)$ of opposite chirality. The global chiral symmetry is spontanously broken down by N mass parameters $ m_{i} $ which pair up the fields in N massive Dirac
fermions, leaving a single massless chiral component or zero mode. Let us signify $U(1)_{Rj}$ and $U(1)_{Li}$ the Abelian factors under which $N +1$ chiral fermions $\psi_{Rj}$ and  N fermions $\psi_{Li}(i = 0, . . . , N-1)$ have charge 1, respectively. Then $m_{j}$ get charges $(1, -1)$ under $U(1)_{Lj} \times U(1)_{Rj}$ , and
$mq_{j}$ get charges $(1, -1)$ under $U(1)_{Lj} \times U(1)_{Rj}$ . One abelian factor of the chiral symmetry or the zero mode in the $\psi_{R}$ sector remains unbroken by both $m$ and $mq$.
\par 
Under the Standard Model gauge group, we expand the Standard Model with $n$ left-handed and $n + 1$ right-handed chiral fermions, singlets, that we represent as $\psi_{Li} (i = 0, ..., n-1)$ and $\psi_{Ri}(i = 0, ..., n)$
respectively. The Lagrangian of the model interprets as:
\begin{equation}
\textit{L}= \textit{L}_{SM} + \textit{L}_{CLOCKWORK} + \textit{L}_{INT}
\end{equation}.
Here $\textit{L}_{SM}$ is the Standard Model Lagrangian, $\textit{L}_{CLOCKWORK}$ is the part of the Lagrangian involving only the new right-handed chiral fermions, singlets, and $ \textit{L}_{INT} $ is the interaction term of the new fields, right-handed chiral fermions, singlets with the Standard Model fields. The Standard Model fields only share coupling to the last one of the fermionic clockwork, right-handed chiral fermion.
\begin{equation}
\textit{L}_{INT} = -Y\tilde{H}\bar{L_{l}}\psi_{Rn}
\end{equation},
where
$\tilde{H}= i\tau_{2}H^{*}$. H is the Standard Model Higgs doublet and $ L_{l} $ is the 
Standard Model left handed leptonic fields.
\par 
The Lagrangian for the fermion fields is
\begin{equation}
\textit{L} = \textit{L}_{kin} - \sum^{n-1}_{j=0}(m_{i}\bar{\psi}_{\textit{Lj}} {\psi}_{\textit{Rj}}- m_{i}^{'}\bar{\psi}_{\textit{Lj}}{\psi}_{\textit{Rj+1}}+ h.c)- \sum^{n-1}_{j=0}(\frac{1}{2}M_{Li}\bar{\psi^{c}_{Li}}\psi_{Li}) - \sum^{n}_{j=0}(\frac{1}{2}M_{Ri}\bar{\psi^{c}_{Ri}}\psi_{Ri}) \equiv \textit{L}_{kin} -(\bar{\psi}_{\textit{L}}M_{ {\psi}} {\psi}_{\textit{R}}+ h.c) 
\end{equation}
where $ \textit{L}_{kin} $ is the kinetic term for all fermions. We take universal values for $m$ and $q$. $m$ is real with a chiral rotation of the fermions, the parameters q are in general complex, but we choose them as real for simplicity. $M$ is a ($2n$ + 1) × ($2n$ + 1) mass matrix. $ \textit{L}_{kin} $ is symmetric with respect to the global gauge group $U(n)_{L} \times U(n + 1)_{R}$. The mass terms $ m_{i} $ break the global symmetry $U(n)_{L} \times U(n + 1)_{R}\rightarrow \prod^{n-1}_{i=0}U(1)_{i} $, where
$U(1)_{i}$ acts as $\psi_{L,i} \rightarrow e^{i \beta_{i}}\psi_{L,i} $, $\psi_{L,i} \rightarrow e^{i \beta_{i}}\psi_{L,i} $, which when combined with the mass terms $ m_{i} $, automatically breaks the global symmetry $U(n)_{L} \times U(n + 1)_{R}\rightarrow U(1)_{CW}$, where under $U(1)_{CW}$ clockwork fermion fields tansform as $\psi_{L,i} \rightarrow e^{i \beta}\psi_{L,i} $, $\psi_{L,i} \rightarrow e^{i \beta}\psi_{L,i} $ for all i. $M_{Li}$ and $M_{Ri}$ are the Majorana masses for the left and right handed singlet fields. For simplicity we assume universal Dirac masses, Majorana masses and nearest neighbor interactions as $m_{i}= m, m_{i}^{'}= mq $, $ M_{Li} = M_{Ri} = m\tilde{q}$ for all i. This conjecture leads to the mass matrix:
\begin{equation}
 M_{\psi} = m\begin{pmatrix}
\tilde{q}& 0&...  & 0 &1 & -q &..  & 0\\
0 & \tilde{q}& ...  & 0 & 0 & 1& ...  & 0\\
&...........................................\\
0 & 0 & ........&\tilde{q} & 0 & 0 & 0 & -q\\
1& 0&.. ........& 0 & \tilde{q} & 0& ...  & 0\\
-\tilde{q}& 1 &...& 0 & 0 & \tilde{q}&......&0\\
&..............................................\\
&...............................................\\
0 & 0 & .......&-\tilde{q} & 0 & 0 & 0 & \tilde{q}\\

\end{pmatrix}.
\end{equation},
The eigen values of the mass matrix are:
\begin{equation}
M_{0}= m\tilde{q}
\end{equation},
\begin{equation}
M_{k}= m\tilde{q} - m\sqrt{\kappa_{r}}, \hspace{0.1cm} r=1......n
\end{equation},

\begin{equation}
M_{n+k}= m\tilde{q} + m\sqrt{\kappa_{r}}, \hspace{0.1cm} r=1......n
\end{equation},
where $ \kappa_{r} $ is defined as
\begin{equation}
\kappa_{r} = q^{2} +1-2q cos\frac{r\pi}{n+1}
\end{equation}.
Some studies on different aspects of Lepton Flavour Violation has been extensively done in {\color{blue}\cite{GG}}. Also $ A_{4} $ symmetry which gives the corrections to the TBM form for the leading order neutrino mixing matrix is studied in {\color{blue}\cite{GG1}}. The predictions of vanishing $ \theta_{13} $ by TBM is owing to its invariance under $ \mu-\tau $ exchange symmetry {\color{blue}\cite{GG}}. Small explicit breaking of $ \mu-\tau $ symmetry can generate large Dirac CP violating phase in the context of neutrino oscillations {\color{blue}\cite{GC}}.
\section{Lepton Flavour Violation in case of $ M_{Li}, M_{Ri}\neq 0$ }
The Yukawa couplings for the zero mode of the clockwork gears are suppressed, thereby  explains why the neutrino masses are small, but the Yukawa couplings for the higher modes of the clockwork gears are unsuppressed which leads to observable effects like  lepton flavor violation at low energies. We find in this work that if the clockwork gear masses are low then signatures of lepton flavor violation existing in the Yukawa couplings of the higher modes through quantum effects generates rare leptonic decays such as$l_{i} → l_{j}+ \gamma$ or $\mu-e$ conversion processes in nuclei, induced by clockwork fermions, with decay rates which could be accessible at next run of LHC.
\par 
We calculate the rate $ l_{i} \rightarrow l_{j} + \gamma $ from {\color{blue}\cite{a}}. For N clockwork generations, we have
\begin{equation}
B\left(\mu \rightarrow e \gamma\right) \simeq \frac{3\alpha_{em}\upsilon^{4}}{8\pi}|\sum^{N}_{\alpha=1}\sum^{n_{\alpha}}_{k=1}\frac{Y_{k}^{e\alpha}Y_{k}^{\mu \alpha}}{M_{k}^{\alpha^{2}}}F \left( x_{k}^{\alpha} \right)  |^{2}
\end{equation}
where $ \alpha_{em} $ signifies the fine stucture constant. $ n_{\alpha} $is the number of clockwork gears in the $ \alpha $ $ - $th generation. $ M_{k}^{\alpha}$ is the mass of the $k-$th mode in the $ \alpha $ $ - $th generation 
$\left( k = 1, ..., n_{\alpha} \right) $. Here
\begin{eqnarray}
x_{k}^{\alpha} \equiv \frac{M_{k}^{\alpha^{2}}}{M_{W}^{2}}
\end{eqnarray}
We define the loop function F(x) as 
\begin{equation}
F(x) \equiv \frac{1}{6 \left( 1-x \right) ^{4}}\left( 10-43x + 78x^{2} -49x^{3} + 4x^{4} - 18x^{3}log x\right) 
\end{equation},
which has limits $ F(0) = \frac{5}{3} $ and $ F(\propto) = \frac{2}{3} $
\par 
For N clockwork generations of clockwork gears we calculate the rate for $ l_{i} \rightarrow l_{j} + \gamma $ from {\color{blue}\cite{1,2,3}}.

\begin{center}
\begin{figure*}[htbp]
\includegraphics[height=6cm,width=8cm]{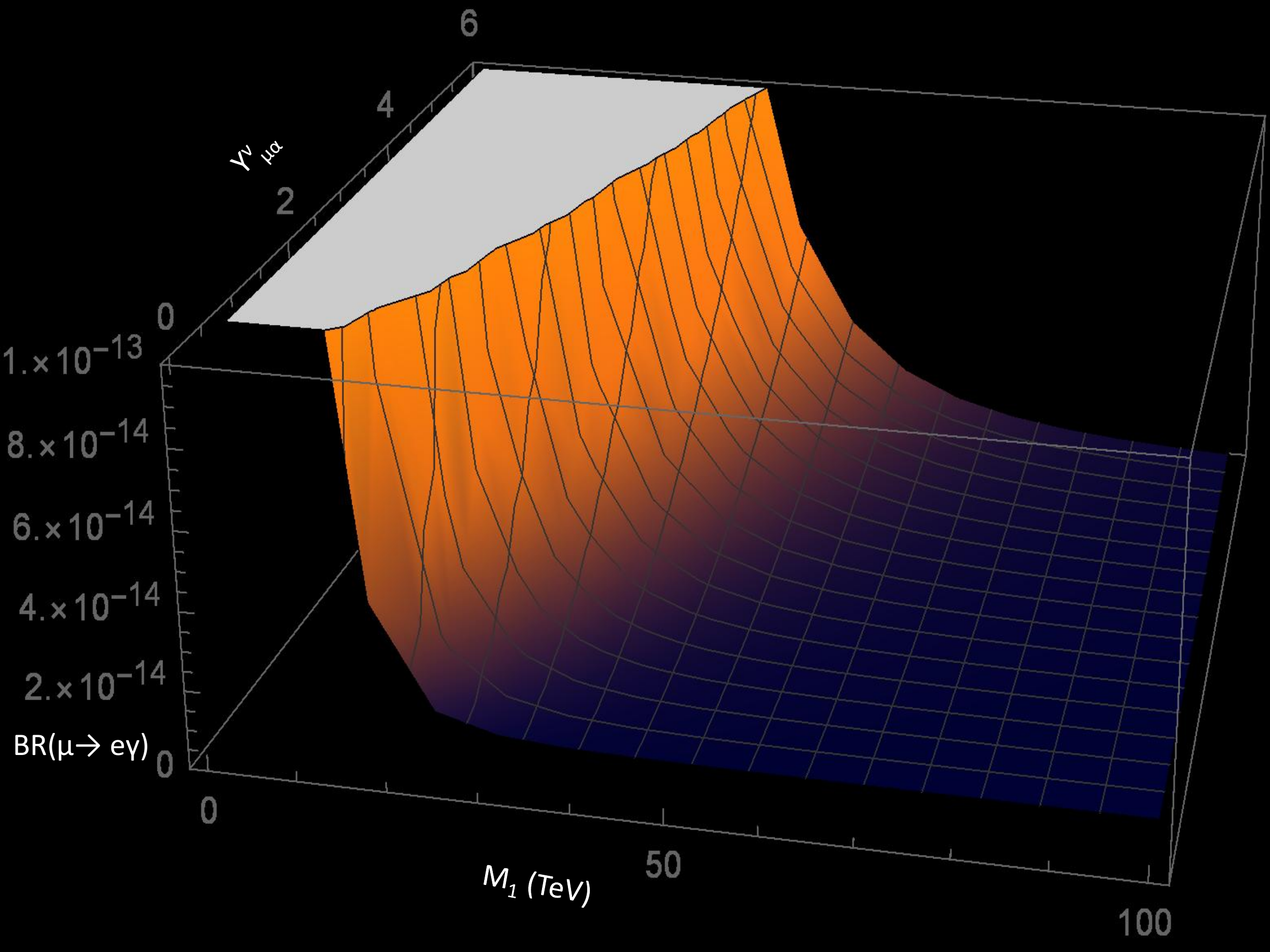}

\caption{Predicted value of $BR(\mu \rightarrow e + \gamma)$ for points of the parameter space reproducing the observed neutrino
oscillation parameters, as a function of the mass of the first clockwork gear and the Yukawas satisfying the charged lepton masses.}
\label{fig:1}
\end{figure*}
\end{center}
\begin{center}

\begin{figure*}[htbp]
\includegraphics[height=9cm,width=13cm]{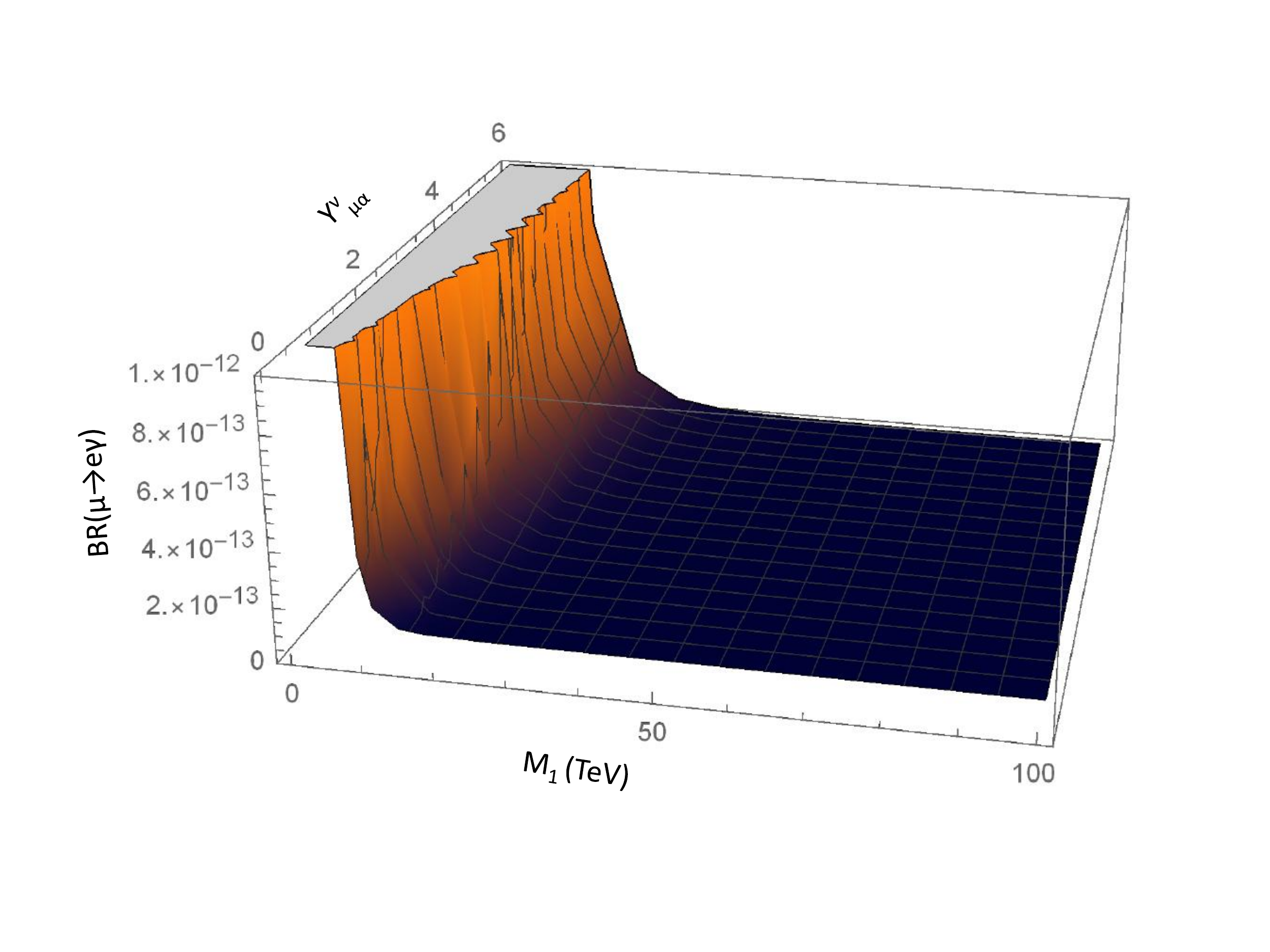}

\caption{Predicted value of $BR(\mu \rightarrow e + \gamma)$ for points of the parameter space reproducing the observed neutrino oscillation parameters as a function of the mass of the first clockwork gear and the Yukawas satisfying the charged lepton masses. It follows from the figure that the clockwork gears must be larger than $ \sim 10 $ TeV in order to evade the experimental constraints, unless very fine cancellations occurs among all contributions to the process $BR(\mu \rightarrow e + \gamma)$.}
\label{fig:1}
\end{figure*}
\end{center}

The present stringent upper bound $BR(\mu \rightarrow e + \gamma) \leq $ $4.2 \times 10^{-13}$ from the MEG experiment poses firm
constraints on the mass scale of the clockwork gears. In {\color{blue}Fig. 1, 2, 3, 4, 5, 6, 7} we present the branching ratio expected for points reproducing the measured neutrino masses as well as charged lepton masses, assuming two clockwork generations, as earned in the figures presented in section 3, as a function of the mass of the first clockwork gear. It is observed from {\color{blue}Fig.1, 2} that if the yukawa modes $Y_{\mu\alpha}^{\nu}$ are of the order $0$ to $6$, then the clockwork gears must be larger than $10$ TeV then the clockwork gears must be larger than $10$ TeV in order to elude the experimental constraints on $BR(\mu \rightarrow e + \gamma)$ as apparent from {\color{blue}Fig.1, 2}, except very fine cancellations occurs among all contributions to this process. Similarly for observation of $BR(\tau \rightarrow e + \gamma)$ for a large number of clockwork generations we look forward to even more restricted lower limits on the lightest gear mass, due to more number of particles in the loop. As evident from {\color{blue}Fig.3, 4}, it follows from the figure that the clockwork gears must be larger than $ \sim 0.2 $ TeV and must be less than 1 TeV for Yukawas corresponding to 0.5 as restricted by the experimental constraints on $BR(\tau \rightarrow e + \gamma) \leq 3.3 \times 10^{-8}$, unless very fine cancellations occurs among all contributions to the process $BR(\tau \rightarrow e + \gamma)$. Also for Yukawas resembling the values around 2.42338 the values of the mass of the first clockwork gear must be less than 2.5 TeV and greater than 0.2 TeV, if the yukawa coupling modes $Y_{\tau\alpha}^{\nu}$ are of the order $0.4$ to $2.4238$. It follows from {\color{blue}Fig.5}, that the clockwork gears must be larger than $ 2 $ TeV and must be less than 5 TeV as restricted by the experimental constraints $BR(\tau \rightarrow \mu + \gamma) \leq 4.4 \times 10^{-8}$,for Yukawa $ Y_{\nu}^{\tau \alpha} $ corresponding to 1 unless very fine cancellations occurs among all contributions to the process $BR(\tau \rightarrow \mu + \gamma)$ and the clockwork gears must be larger than $ \sim 2 $ TeV and must be less than $40$ TeV restricted by the experimental constraints $BR(\tau \rightarrow \mu + \gamma) \leq 4.4 \times 10^{-8}$, for Yukawa $ Y_{\nu}^{\tau \alpha} $ corresponding to 7.67. It is found from {\color{blue}Fig.7} that the clockwork gears must be larger than $ \sim 90 $ TeV and must be less than 200 TeV as constained by the experimental upper bound on  $BR(\tau \rightarrow e + \gamma) \leq 4.4 \times 10^{-8}$,for Yukawa $ Y_{\nu}^{\tau \alpha} $ corresponding to $1$ to $2$ unless very fine cancellations occurs among all contributions to the process $BR(\tau \rightarrow \mu + \gamma)$ and the clockwork gears must be greatee than $ \sim 90 $ TeV for Yukawa $ Y_{\nu}^{\tau \alpha} $ corresponding to $2$ to $7.67$ as constrained by experimental limit.  
\begin{center}
\begin{figure*}[htbp]
\includegraphics[height=9cm,width=13cm]{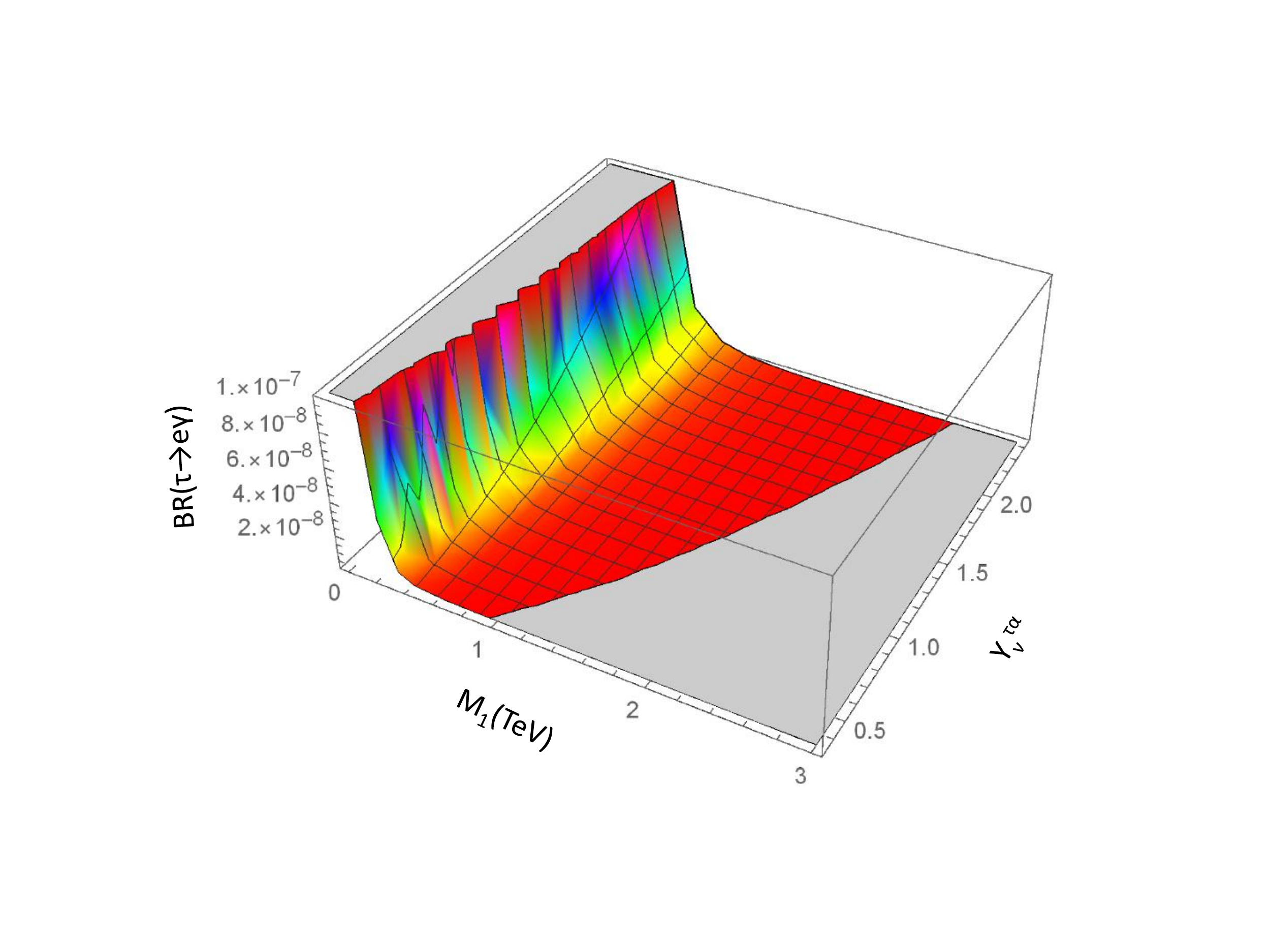}
\caption{Predicted value of $BR(\tau \rightarrow e + \gamma)$ for points of the parameter space reproducing the observed neutrino
oscillation parameters, as a function of the mass of the first clockwork gear. It follows from the figure that the clockwork gears must be larger than $ \sim 0.2 $ TeV and must be less than 1 TeV for Yukawas corresponding to 0.5 as restricted by the experimental constraints on $BR(\tau \rightarrow e + \gamma) \leq 3.3 \times 10^{-8}$, unless very fine cancellations occurs among all contributions to the process $BR(\tau \rightarrow e + \gamma)$. Also for Yukawas resembling the values around 2.42338 the values of the mass of the first clockwork gear must be less than 2.5 TeV and greater than 0.2 TeV. }
\label{fig:1}
\end{figure*}
\end{center}

\begin{center}
\begin{figure*}[htbp]
\includegraphics[height=7.5cm,width=13cm]{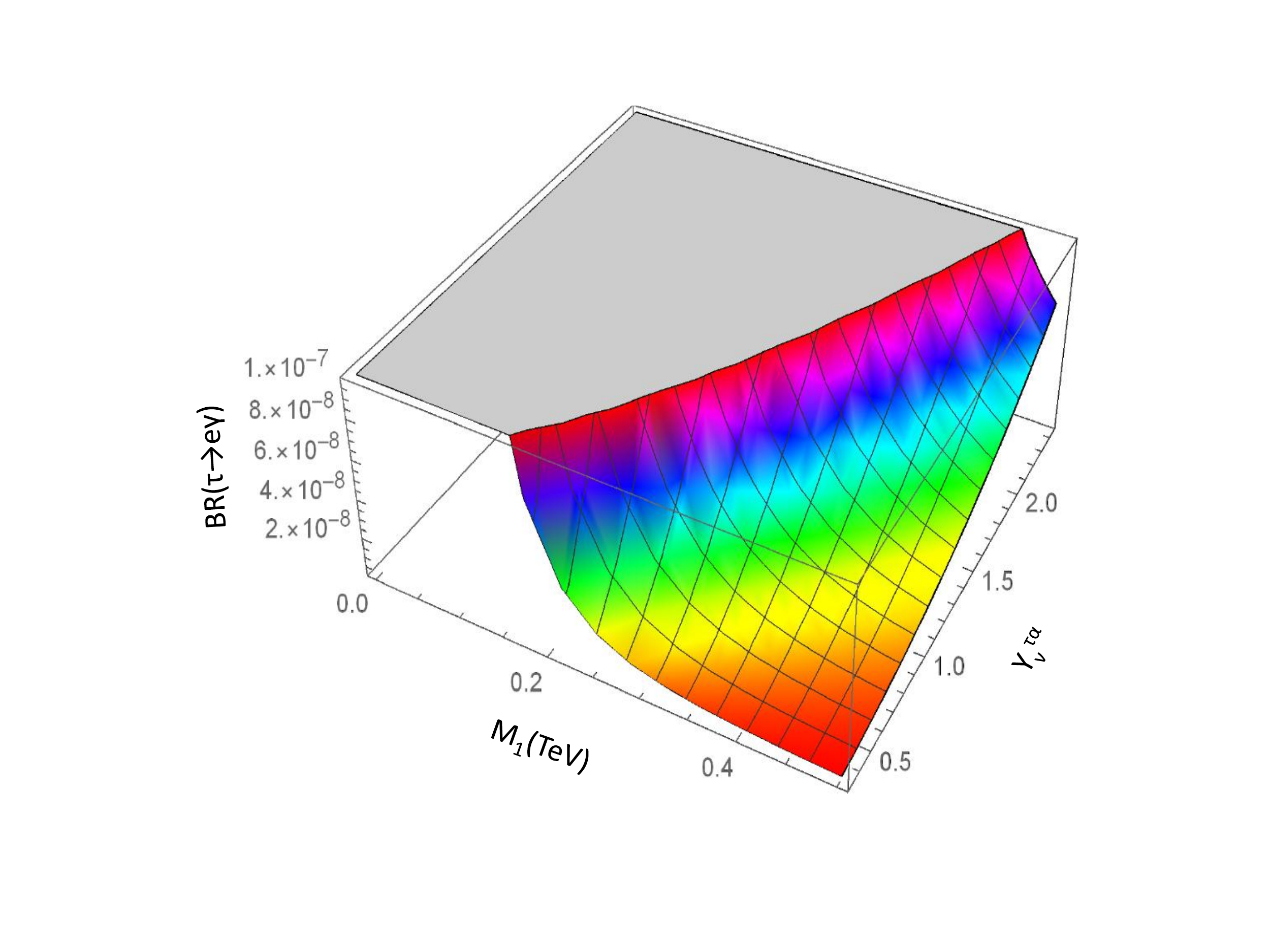}

\caption{Predicted value of $BR(\tau \rightarrow e + \gamma)$ for points of the parameter space reproducing the observed neutrino
oscillation parameters, as a function of the mass of the first clockwork gear and the Yukawas satisfying the charged lepton masses.}
\label{fig:1}
\end{figure*}
\end{center}

\begin{center}
\begin{figure*}[htbp]
\includegraphics[height=9cm,width=13cm]{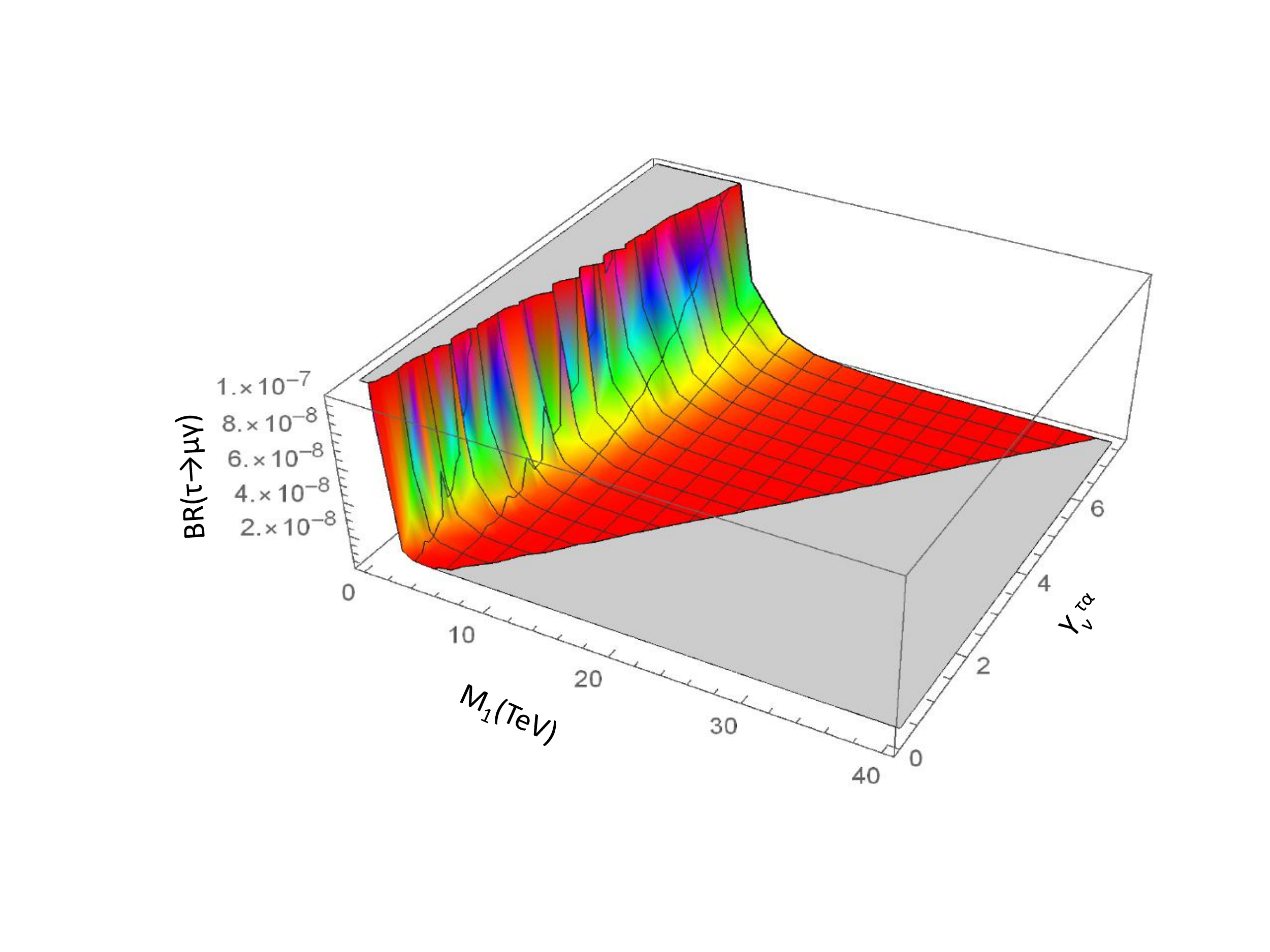}

\caption{Predicted value of $BR(\tau \rightarrow \mu + \gamma)$ for points of the parameter space reproducing the observed neutrino
oscillation parameters, as a function of the mass of the first clockwork gear. It follows from the figure that the clockwork gears must be larger than $ 2 $ TeV and must be less than 5 TeV restricted by the experimental constraints $BR(\tau \rightarrow \mu + \gamma) \leq 4.4 \times 10^{-8}$,for Yukawa $ Y_{\nu}^{\tau \alpha} $ corresponding to 1 unless very fine cancellations occurs among all contributions to the process $BR(\tau \rightarrow \mu + \gamma)$ and the clockwork gears must be larger than $ \sim 2 $ TeV and must be less than $40$ TeV restricted by the experimental constraints $BR(\tau \rightarrow \mu + \gamma) \leq 4.4 \times 10^{-8}$,for Yukawa $ Y_{\nu}^{\tau \alpha} $ corresponding to 7.67.}
\label{fig:1}
\end{figure*}
\end{center}

\begin{center}
\begin{figure*}[htbp]
\includegraphics[height=9cm,width=13cm]{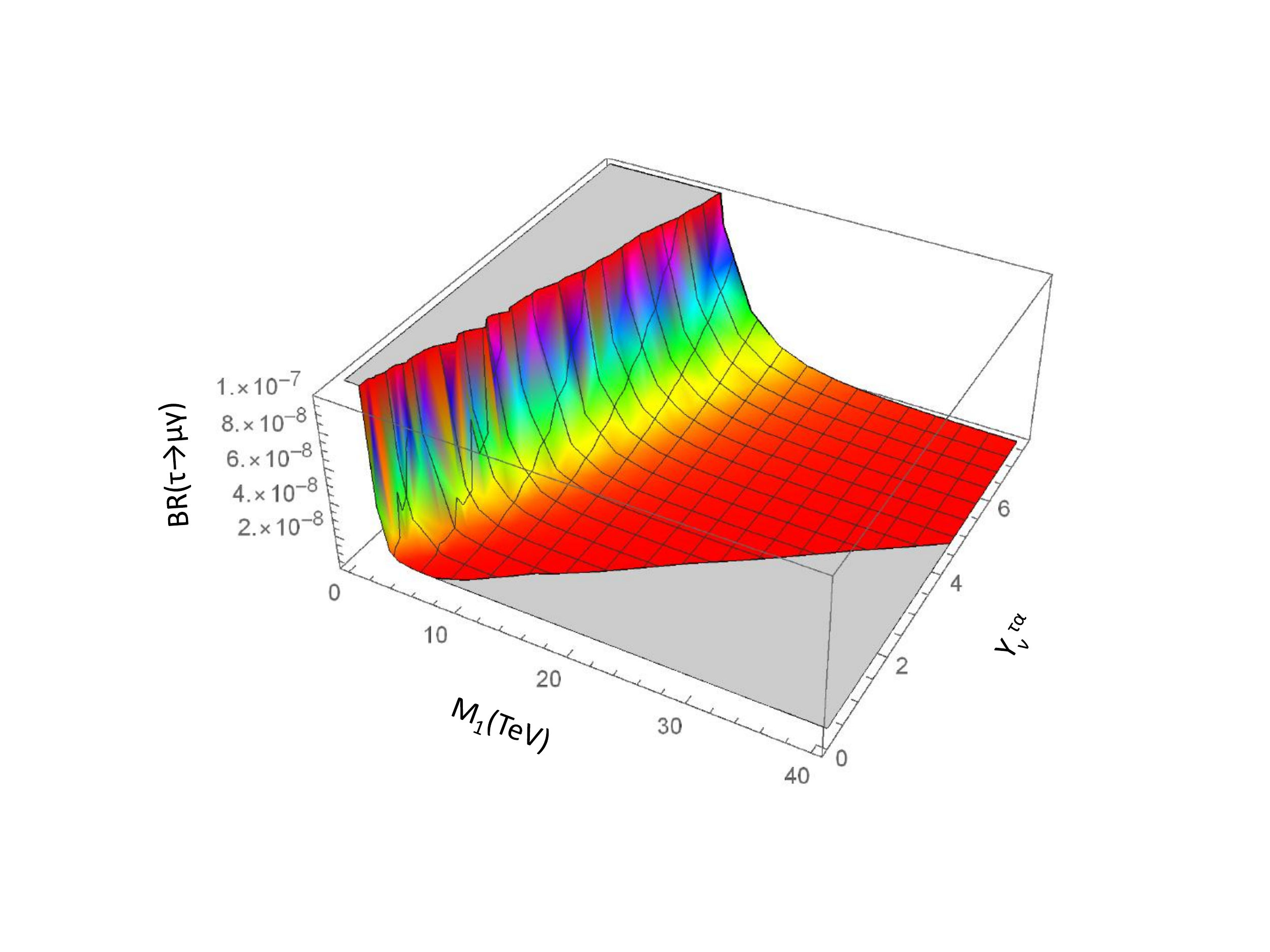}

\caption{Predicted value of $BR(\tau \rightarrow \mu + \gamma)$ for points of the parameter space reproducing the observed neutrino
oscillation parameters, as a function of the mass of the first clockwork gear.}
\label{fig:1}
\end{figure*}
\end{center}

\begin{center}
\begin{figure*}[htbp]
\includegraphics[height=9cm,width=13cm]{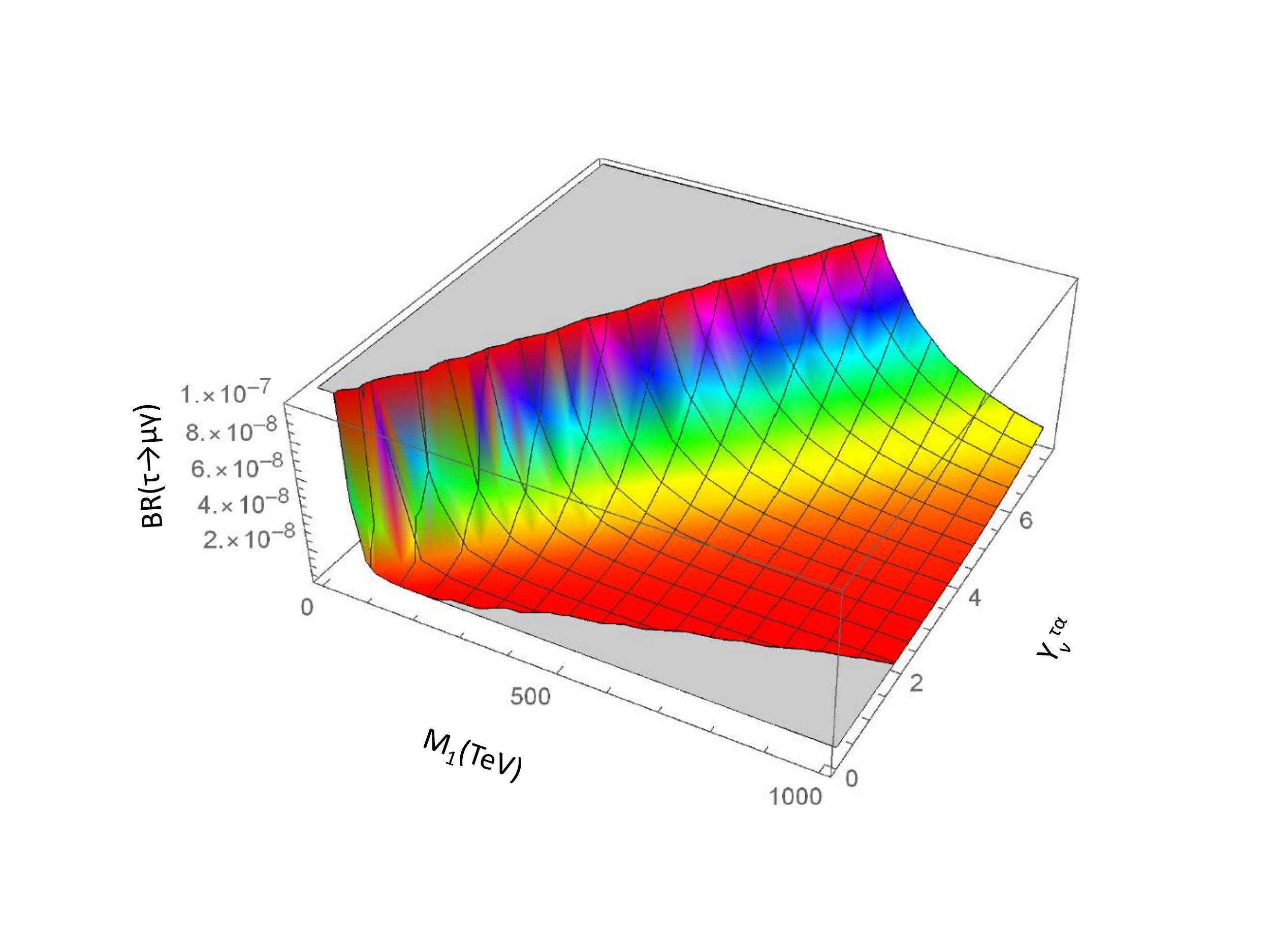}

\caption{Predicted value of $BR(\tau \rightarrow \mu + \gamma)$ for points of the parameter space reproducing the observed neutrino
oscillation parameters, as a function of the mass of the first clockwork gear. It follows from the figure that the clockwork gears must be larger than $ \sim 90 $ TeV and must be less than 200 TeV as constained by the experimental upper bound on  $BR(\tau \rightarrow e + \gamma) \leq 4.4 \times 10^{-8}$,for Yukawa $ Y_{\nu}^{\tau \alpha} $ corresponding to $1$ to $2$ unless very fine cancellations occurs among all contributions to the process $BR(\tau \rightarrow \mu + \gamma)$ and the clockwork gears must be greatee than $ \sim 90 $ TeV for Yukawa $ Y_{\nu}^{\tau \alpha} $ corresponding to $2$ to $7.67$ as constrained by experimental limit.}
\label{fig:1}
\end{figure*}
\end{center}

\section{Clockwork Photon}
A clockwork photon can be acquired by clockworking the invariance of gauge symmetry. Let us take $ N+1$ $ U(1)$ gauge groups with same gauge coupling $g$, and N complex scalars, $ \psi_{j} $, where $j = 0,......,N-1$ having charge $\left(1,-q \right )$. The underlying gauge symmetry is $U(1)_{j} \times U(1)_{j+1}$. Given these scalars acquire 
a negative mass$-$squared, which generates vacuum expectation values which are
at the same scale $\lambda$ . The Lagrangian is
\begin{equation}
\textit{L} = -\sum^{N}_{j=0}\frac{1}{4}F_{\alpha\beta}^{j}F^{j\alpha\beta}-\sum^{N-1}_{j=0}[|D_{\alpha}\psi_{j}|^{2}+\lambda(|\psi_{j}|^{2}-\frac{1}{2}\lambda^{2})]
\end{equation}
where
\begin{equation}
D_{\alpha}\psi_{j} = [\delta_{\alpha}-ig(A_{\alpha}^{j}-qA_{\alpha}^{j+1})]\psi_{j}
\end{equation}
The spontaneous symmetry breaking is of the form $U(1)^{N+1} \rightarrow U(1)$
\begin{equation}
\textit{L} = -\sum^{N}_{j=0}\frac{1}{4}F_{\alpha\beta}^{j}F^{j\alpha\beta}-\sum^{N-1}_{j=0}\frac{g^{2}f^{2}}{2}(A_{\alpha}^{j}-qA_{\alpha}^{j+1})^{2}
\end{equation}
The above implicit Lagrangian is below the scale $ \lambda $ while working in unitary gauge field. The mass matrix is same as of the clockwork form, where the heavy gauge bosons are the photon gear $ A_{\alpha} $ and one photon remaining massless. The clockwork photon has important implications. When matter gets charge only under the Abelian factor identical to the last site of the clockwork gear, then the clockwork phenomenon will produce exponentially small couplings to the massless photon  which generates visible particles with millicharges. Heavy photon gears could be attainable in next run of collider searches.
\section{Clockwork Graviton}
For $N+1$ copies of general relativity, there are $N+1$ analogous massless gravitons.
Each graviton can be descrribed through an extension of the metric around flat space$-$time,
\begin{equation}
g^{\mu\nu}_{i} = \varepsilon^{\mu\nu}_{i} + \frac{2}{M^{2}_{i}}h^{\mu\nu}_{i}
\end{equation}
The clockwork mechanism breaks $N + 1$ copies of diffeomorphism invariance to one diffeomorphism invariance at the linear level through near-neighbour Pauli-Fierz term attire for massive gravitons.
\begin{equation}
\textit{L} = -\frac{m^{2}}{2}\sum^{N-1}_{i=0}([h^{\mu\nu}_{i} -qh^{\mu\nu}_{i+1} ]-[\varepsilon_{\mu\nu}(h^{\mu\nu}_{i} -qh^{\mu\nu}_{i+1})]^{2})
\end{equation}
The massive lagrangian in Eq.(16) is invariant under the gauge symmetry
\begin{equation}
h^{\mu\nu}_{i} \rightarrow h^{\mu\nu}_{i} + \frac{1}{q^{i}}(\delta^{\mu}A^{\nu}+\delta^{\nu}A^{\mu})
\end{equation}
where $A^{\mu}$ is a space-time vector. This gauge symmetry imposes the masslessness of the clockwork graviton and is maintained by the clockwork form of the mass terms. The
mass matrix is of the clockwork structure, with the massive gravitons being the gears and 
one remaining as zero mass graviton. If the energy$-$momentum tensor $ T^{\mu\nu} $  couples only to the last site of the clockwork gear at Planck alike scale $M_{N}$, then the coupling to the zero mass graviton is,
\begin{equation}
-\frac{1}{M_{N}}h^{\mu\nu}_{N}T_{\mu\nu}\rightarrow-\frac{1}{M_{P}}\tilde{h}^{\mu\nu}_{0}T_{\mu\nu},
\end{equation}
where, $M_{P} = \frac{q^{N}M^{N}}{N_{0}}$. The effective Planck scale $M_{P}$, that realises the strength of gravity in the low-energy scale of the theory, is exponentially larger than $M_{N}$ amplified by a factor $q_{N}$ . This provides the clockwork mechanism a result to the hierarchy problem in where new physics, like quantum gravity, sets near the weak scale, in total resemblance as afforded by Large Extra Dimensions {\color{blue}\cite{x}} and the warped extra dimensional model of Randall$-$Sundrum {\color{blue}\cite{y}}. The metric of the clockwork space$-$time is similar to a 5D metric known as the linear dilaton model which resembles the dual of Little String Theory {\color{blue}\cite{z}}, after compressing additional dimensions. This counts for a signature of hint towards an untrodden connection between string theory and the continuum clockwork theory, as it distinguishes the clockwork solution of the hierarchy nature with the Little String Theory solution. 
\section{Conclusion}
The clockwork is a significant mechanism for promoting exponentially suppressed
interactions within a quantum field theory containing only $O(1)$ parameters, in natural units, and a definite number of clockwork fields. As a result, one can stir up exponentially large interaction scales, even when no new physics materializes at this high energy scale. Clockwork holds in models of inflation or relaxation to prompt apparent super Planckian field excursions. 

We have scutinised in detail the impact of the Majorana masses in the clockwork sector in the generation of lepton favour violation. When the Majorana masses are
non-vanishing, the zero mode of the clockwork gear is no longer massless. But, the corresponding Yukawa coupling still inspite of that has the clockwork structure. In this case, small neutrino masses arises of the interaction between the standard seesaw mechanism and the “clockworked” Yukawa couplings, and needs very large Majorana masses  to generate the small neutrino mass scale inferred as concluded from oscillation
experiments. The Standard Model leptons couple to the fermions of the clockwork sector with a site dependent field strength, which prompts lepton flavour violating charged current, neutral current and Higgs boson interactions. From the non findings of
the rare leptonic decay scale of the clockwork fermions due to the non-observation of the rare cLFV decay $ \mu\rightarrow e\gamma $, $ \tau \rightarrow \mu \gamma $, $ \tau\rightarrow e\gamma $ our results stir up that the lightest particle of the clockwork sector must have a mass greater than 10 TeV, 0.2 TeV to 2.5 TeV and 2 TeV to 40 TeV respectively if the Yukawa couplings of exponentially suppressed interactions within a quantum field theory contains only $O(1)$ parameters.

\section{Acknowledgement}
GG would like to thank would like to thank University Grants Commission RUSA, MHRD, Government of India for financial support. GG would also like to thank Prof. Sudhir Vempati for useful discussion on this topic.

\end{document}